\documentclass{raa_twocolumn}
\usepackage{graphicx,times}
\usepackage{natbib}
\usepackage{amssymb,amsmath}
\usepackage[justification=centering]{caption}
\usepackage{xspace}
\usepackage{booktabs}
\newcommand{\target}{CAPERS-39810\xspace}

\bibpunct{(}{)}{;}{a}{}{,}
\newcommand{\oiii}{[O\,\textsc{iii}]}
\newcommand{\hei}{He\,\textsc{i}}
\newcommand{\ha}{H$\alpha$}
\newcommand{\hb}{H$\beta$}

\usepackage[pagebackref=true]{hyperref}

\begin{document}

   \title{Discovery of an Extremely Metal-Poor Galaxy at $z=3.654$ Using JWST Infrared Spectroscopy}

 \volnopage{ {\bf 20XX} Vol.\ {\bf X} No. {\bf XX}, 000--000}
   \setcounter{page}{1}

   \author{Sijia Cai  
   \inst{1,*}\footnotetext{$*$Corresponding Authors.}, Zijian Yu\inst{2}
   }

   \institute{ Department of Astronomy, Tsinghua University, Beijing 100084, China; {\it caisj23@mails.tsinghua.edu.cn}\\
        \and
             Blair Academy, 2 Park St, Blairstown, New Jersey 07825\\
\vs \no
   {\small Received 20XX Month Day; accepted 20XX Month Day}
}

\abstract{We report the discovery of an extremely metal-poor galaxy at a redshift of $z = 3.654$, identified through infrared spectroscopy using the James Webb Space Telescope (JWST). This galaxy, CAPERS-39810, exhibits a metallicity of $12 + \log(\mathrm{O/H}) = 6.73\pm0.13$, indicative of its primitive chemical composition, resembling the early stages of galaxy formation in the Universe. We use JWST NIRSpec/MSA for spectroscopic analysis, complemented by photometric data from the COSMOS2025 catalog. Our analysis employs the R3 strong-line diagnostic method to estimate metallicity, due to the lack of auroral lines in the spectrum. The galaxy's emission lines, including \hb, \oiii, \ha~and \hei, are clearly detected. The rest-frame equivalent widths of the strong hydrogen recombination lines are EW$_0$(\hb) = $184\pm48$ \AA~and EW$_0$(\ha) = $1144\pm48$ \AA. Furthermore, we perform detailed spectral energy distribution modeling to derive a galaxy logarithmic stellar mass of $8.02^{+0.22}_{-0.34}$ $M_\odot$. This discovery adds to the growing body of evidence for the existence of very low-metallicity galaxies existed at cosmic noon of $z\approx3$, which are crucial for understanding the processes of chemical enrichment and star formation in young galaxies at the cosmic noon.
}

   \authorrunning{Z.-J. Yu \& S.-J. Cai }            
   \titlerunning{An extremely metal-poor galaxy at $z = 3.654$}  
   \maketitle

\renewcommand\thefootnote{}
\footnotetext{The authors contributed equally to this work and are listed in alphabetical order.}

%
\section{Introduction}           
\label{sect:intro}

The first generations of stars, commonly referred to as Population~III (Pop~III) stars, are thought to have formed out of metal-free primordial gas in the early Universe \citep{Tumlinson2000ApJ...528L..65T,Schaerer2003A&A...397..527S,Bromm2011ARA&A..49..373B}. Owing to the lack of metal-line cooling, Pop~III stars are predicted to reach very high stellar masses and to host extremely hard ionizing radiation fields \citep{Klessen2023ARA&A..61...65K,Nandal2025ApJ...994L..11N}. As a consequence, observational searches have frequently focused on nebular signatures associated with such hard spectra, such as strong He\,\textsc{ii} $\lambda1640$ emission and the absence or weakness of metal lines like [O\,\textsc{iii}] \citep[e.g.,][]{Cai2011ApJ...736L..28C,Cai2015ApJ...799L..19C,Wang2024ApJ...967L..42W,Maiolino2024A&A...687A..67M,Mondal2025ApJ...988..171M,Nakajima2025arXiv250611846N,Fujimoto2025arXiv251211790F}. However, very massive stars evolve rapidly, and the He\,\textsc{ii} emission powered by Pop~III stars is expected to fade on timescales of $\sim$3~Myr \citep{Schaerer2002A&A...382...28S}. Moreover, recent models predict that Pop~III galaxies are likely to be rapidly ``self-polluted" by the first supernova explosions \citep{Rusta2025ApJ...989L..32R}, yielding \oiii/\hb~$\approx 1$. As a result, the unambiguous identification of Pop~III stellar populations remains a major open challenge for both observations and theory.

The James Webb Space Telescope (JWST) has enabled unprecedented observations of faint galaxies in the early Universe, providing new insights into the history of early star formation. Extensive JWST surveys of dwarf galaxies at $z \sim 2\text{-}10$ have revealed a well-defined mass-metallicity relation (MZR), with galaxies at fixed stellar mass exhibiting progressively lower metallicities toward higher redshift \citep{Li2023ApJ...955L..18L,Nakajima2023ApJS..269...33N,Curti2024A&A...684A..75C,Sanders2024ApJ...962...24S,Sarkar2025ApJ...978..136S}. Notably, the metallicities inferred for these systems remain above $\sim 2\%Z_\odot$, a result that has been interpreted either as evidence for rapid chemical enrichment in the early Universe \citep{Gutcke2022ApJ...941..120G} or as a consequence of observational selection biases. At the same time, however, an increasing number of recent studies have reported candidates for extremely metal-poor galaxies (EMPGs), suggesting that this apparent ``metallicity floor" may not be universal \citep{Chemerynska2024ApJ...976L..15C,Hsiao2025arXiv250503873H,Willott2025ApJ...988...26W,Cullen2025MNRAS.540.2176C,Nakajima2025arXiv250611846N}. 
In particular, several recent discoveries have highlighted compelling candidates for primordial or Pop~III-hosting systems. For example, \citet{Morishita2025arXiv250710521M} identified the ``AMORE6" at $z = 5.725$ using NIRCam Wide Field Slitless Spectroscopy (WFSS). At somewhat lower redshifts, closer to the peak epoch of cosmic star formation, \citet{Cai2025ApJ...993L..52C} and \citet{Vanzella2026A&A...705L..12V} reported the candidate primordial galaxies ``CR3" at $z = 3.19$ and ``LAP2" at $z = 4.19$, respectively, based on deep NIRSpec spectroscopy. Together, these discoveries further complicate our understanding of the chemical enrichment history of the Universe.

In this work, we report the discovery of a extreme metal-poor galaxy at z=3.654 and analyze its spectroscopy and spectral energy distributio (SED) to assess metallicity and physical conditions, placing it among the growing population of chemically primitive systems that JWST has so far uncovered. This paper is organized as follows: In Sec. \ref{sec:obs}, we describe our observations, including data selection and processing. In Sec. \ref{sec:method}, we present our procedures for emission line flux measurements, metallicity calibrations based on strong-line diagnostics and SED fitting. Discussion and a summary of our major conclusions are detailed in Sec. \ref{sec:results}. Throughout this work we adopt a flat $\Lambda$CDM cosmology consistent with \citet{Planck2020A&A...641A...6P}, with $H_0 = 67.7~\mathrm{km\,s^{-1}\,Mpc^{-1}}$, $\Omega_{\rm m} = 0.31$, and $\Omega_\Lambda = 0.69$. For the reference gas-phase metallicity scale, we assume a solar oxygen abundance of $12+\log(\mathrm{O/H}) = 8.69$ from \citet{Asplund2021A&A...653A.141A}.

\begin{figure*}
    \centering
    \includegraphics[width=\linewidth]{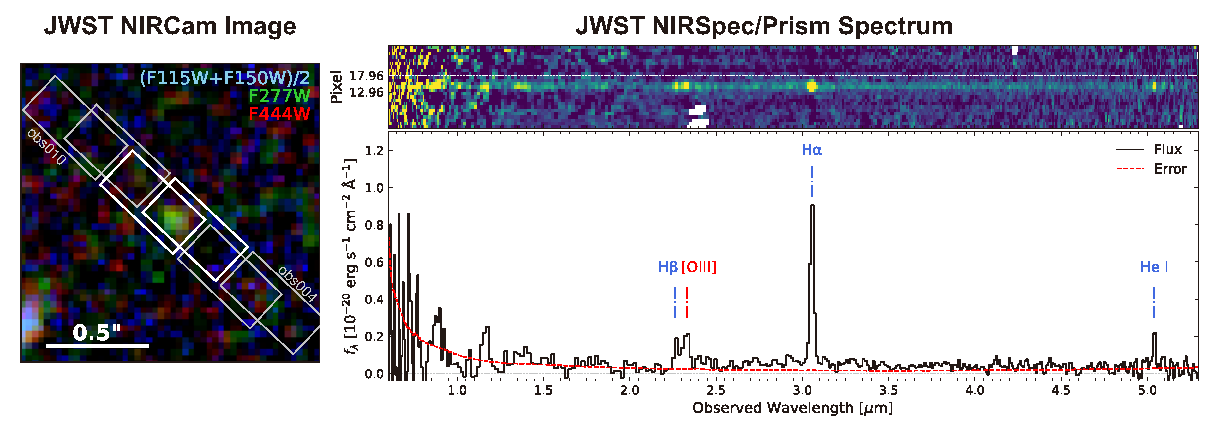}
    \caption{JWST observations of \target are shown. The left panel displays a composite RGB image constructed from JWST NIRCam imaging, where the F444W, F277W, and the average of the F115W and F150W bands are mapped to the red, green, and blue channels, respectively. The white polygon denotes the configuration of the NIRSpec shutters. The right panel shows the stacked NIRSpec/Prism 2D spectrum together with the extracted 1D spectrum. The spectral extraction region is indicated by the white dashed lines. The black solid line and the red dashed line correspond to the measured flux and the associated uncertainty. Key emission features, including \hb, \ha, \hei\,$\lambda$10830, and the metal line \oiii\,$\lambda\lambda$4959, 5007, are identified.}
    \label{fig:spec}
\end{figure*}

\section{Observations and Data Processing}
\label{sec:obs}

\target is located in the well-known COSMOS field \citep{Scoville2007ApJS..172....1S,Koekemoer2011ApJS..197...36K,Grogin2011ApJS..197...35G} at RA = 150.1335857 and Dec = 2.271014. In this work, we primarily use imaging and spectroscopic data obtained with the JWST.

\subsection{Spectroscopy}
The NIRSpec/MSA spectrum of \target was acquired under the JWST Cycle 3 program CANDELS-Area Prism Epoch of Reionization Survey (CAPERS; GO\#6368; PI: M. Dickinson). The main goal of CAPERS is to identify up to 10,000 spectroscopically confirmed galaxies at z $>$ 10 using the NIRSpec Prism. \target was observed in Observation 10 on 15 April 2025 and Observation 4 on 19 May 2025, with 4.74 hours of effective exposure per visit (9.48 hours in total). The raw data are retrieved from the Mikulski Archive for Space Telescopes (MAST) at the Space Telescope Science Institute. Each visit is processed independently using the official JWST pipeline\footnote{\url{https://github.com/spacetelescope/jwst}} (software version 1.17.1; CRDS context \texttt{jwst\_1322.pmap}; \citealt{bushouse_2025_14597407}). The standard Level 1 and Level 2 processing steps, including the 1/f noise correction, are applied. We perform background subtraction using the standard three-shutter nodding strategy to obtain 2D spectra. Finally, 1D spectra is extracted using a box aperture with a width of six pixels, adopting the default extraction parameters. Prior to stacking, the 2D spectra are aligned along the spatial direction to match the trace position in Observation 4, and the spectral axis of both the 1D and 2D products are resampled onto the wavelength grid of Observation 4. The resampled spectra are then combined using inverse-variance weighting to produce the final stacked 1D and 2D spectra showed in Fig. \ref{fig:spec}.

\subsection{Image and Photometry}

We use publicly available imaging and photometric data from the COSMOS2025 catalog \citep{Shuntov2025A&A...704A.339S}. This includes HST/ACS F814W from the COSMOS HST ACS F814W dataset \citep{Koekemoer2007ApJS..172..196K}, JWST/NIRCam F115W, F150W, F277W, and F444W, and JWST/MIRI F770W from the 255-hour JWST Cycle 1 Treasury Program COSMOS-Web (GO\#1727; PIs: J. Kartaltepe and C. Casey; \citealt{Casey2023ApJ...954...31C}). We use the catalog photometry, which is measured in small elliptical apertures and has been corrected for Kron aperture losses and for point-spread function (PSF) effects.

\section{Methods and Analysis}
\label{sec:method}

\subsection{Emission lines}

As seen in the spectrum (Fig. \ref{fig:spec}), four emission lines are robustly detected at a significance level of greater than 3$\sigma$, including \hb, \oiii\,$\lambda\lambda$4959, 5007, \ha~and \hei\,$\lambda$10830.
Due to limited spectral resolution, the two emission lines \oiii\,$\lambda$4959 and \oiii\,$\lambda$5007 are blended. 
We thus model the combined \oiii\,$\lambda\lambda$4959,5007 profile with a single Gaussian function.
Prior to line fitting, we mask the short-wavelength region (observed wavelength $\sim$0.6-2.0 $\mu$m), where the noise level is high. The continuum and emission lines were modeled simultaneously, with the continuum represented by a constant component and the emission lines described by Gaussian profiles. The best-fit parameters were obtained through non-linear least-squares minimization, from which the line fluxes and equivalent widths (EWs) were derived.

We measure line fluxes of $f(\mathrm{H}\beta) = (3.8\pm1.0) \times 10^{-19}$ erg s$^{-1}$ cm$^{-2}$, $f([\mathrm{OIII}]) = (7.9\pm1.6) \times 10^{-19}$ erg s$^{-1}$ cm$^{-2}$, $f(\mathrm{H}\alpha) = (23.6\pm0.8) \times 10^{-19}$ erg s$^{-1}$ cm$^{-2}$ and $f(\mathrm{HeI}) = (3.1\pm0.8) \times 10^{-19}$ erg s$^{-1}$ cm$^{-2}$, with a rest-frame equivalent width of $\mathrm{EW}_0(\mathrm{H}\beta) = 184 \pm 48$~\AA, $\mathrm{EW}_0(\mathrm{H}\alpha) = 1144 \pm 48$~\AA~and $\mathrm{EW}_0(\mathrm{HeI}) = 152 \pm 38$~\AA. The measured value of all detected emission lines are listed in Table~\ref{tab:emission}.

\subsection{Gas-phase Metallicity}

The most reliable and physically motivated metallicity calibration, the direct T$_\mathrm{e}$ method, requires measurements of key physical conditions, including the electron temperature and electron density. However, temperature-sensitive auroral lines, such as \oiii\,$\lambda$4363, are typically faint, particularly in high-redshift EMPGs. In addition, the \oiii\,$\lambda$3727 line falls on the low-signal-to-noise blue end of the NIRSpec/Prism spectrum. Given these limitations, we instead adopt a strong-line metallicity diagnostic. In our case, the gas-phase metallicity is constrained using the R3 = \oiii\,$\lambda$5007/\hb~ratio. The \oiii\,$\lambda$5007 line flux is estimated by adopting the theoretical \oiii\,$\lambda$5007/\oiii\,$\lambda$4959 ratio of 2.98 \citep{Storey2000MNRAS.312..813S}. 

Following \citet{Sanders2024ApJ...962...24S}, we adopt the R3 extrapolation calibration, in which the gas-phase metallicity is parameterized as a polynomial function of the strong-line ratio. The relation is given by

\begin{equation}
    \log(\mathrm{R3}) = 0.834 - 0.072x - 0.453x^2
\end{equation}

where $x = 12 + \log(\mathrm{O}/\mathrm{H}) - 8.0$. Given the double-valued nature of the R3 calibration, we adopt the low-metallicity branch based on the trend of mass-metallicity relation and the stellar mass derived from SED fitting (Sec. \ref{sec:sed}). We determine the metallicity of CAPERS-39810 to be $12+\log(\mathrm{O}/\mathrm{H}) = 6.73\pm0.13$. 
We did not apply an explicit dust correction to the R3 ratio used for the metallicity estimate. 
\oiii\,$\lambda$5007 and \hb~are closely separated in wavelength and therefore experience nearly identical dust attenuation. 
Even if the reddening implied by the Balmer decrement is adopted and a standard attenuation 
law \citep{Calzetti2000ApJ...533..682C} is applied, the resulting change in the line ratio leads to a shift of only 
$\sim0.05$\,dex in $12+\log(\mathrm{O/H})$, which is well within the uncertainty of the metallicity estimate.

\begin{table}[ht]
\centering\caption{\textbf{Emission Line Properties}}
\label{tab:emission}
\renewcommand{\thefootnote}{\alph{footnote}}

\begin{tabular}{lll}
\toprule\toprule
\textbf{Parameter} & \textbf{Value} & \textbf{Unit} \\
\midrule
\midrule
$f(\mathrm{H}\beta)$ & $(3.8 \pm 1.0)\times 10^{-19}$ & cgs \\

$f([\mathrm{O}\,\text{III}]~\lambda4959,5007)$ & $(7.9 \pm 1.6)\times 10^{-19}$ & cgs\\

$f(\mathrm{H}\alpha)$ & $(23.6 \pm 0.8)\times 10^{-19}$ & cgs \\

$f(\mathrm{He}\,\text{I}~\lambda10830)$ & $(3.1 \pm 0.8)\times 10^{-19}$ & cgs\\

\midrule

$\mathrm{EW}_0(\mathrm{H}\beta)$ & $184 \pm 48$ & \AA \\

$\mathrm{EW}_0(\mathrm{H}\alpha)$ & $1144 \pm 48$ & \AA \\

$\mathrm{EW}_0(\mathrm{He\,I})$ & $152 \pm 38$ & \AA \\

\midrule

$\mathrm{R}3$ & $1.6 \pm 0.5$ & - \\

$12+\log(\mathrm{O}/\mathrm{H})$ & $6.73\pm0.13$ & - \\
\bottomrule\bottomrule
\end{tabular}
\tablecomments{0.9\textwidth}{Flux is given in \text{cgs} units as erg s$^{-1}$ cm$^{-2}$.}
\end{table}

\subsection{SED Modeling}
\label{sec:sed}

We perform SED fitting for our extremely metal-poor galaxy using the Bayesian Analysis of Galaxies for Physical Inference and Parameter EStimation code, \texttt{Bagpipes} \citep{Carnall2018MNRAS.480.4379C}. The SED modeling is based on JWST/NIRCam broad-band photometry and is used to estimate the global physical properties of the galaxy, in particular its stellar mass and approximate stellar age. Adopting photometric SED fitting also avoids the additional complexity and potential systematic uncertainties associated with incorporating low signal-to-noise ratio (SNR) spectroscopic data, while remaining fully adequate for resolving the mass-metallicity degeneracy relevant to our metallicity determination.

To construct the SED model, we adopt a \citet{Kroupa2001MNRAS.322..231K} initial mass function (IMF) and assume a delayed-$\tau$ star formation history (SFH), parameterized as

\[ \mathrm{SFR}(t) \propto t \, \exp\!\left(-\frac{t}{\tau}\right), \] 

which allows for an initial rise followed by an exponential decline in the star formation rate.  We adopt uniform priors on the model parameters, with the stellar age allowed to vary between 0.01 and 1.86 Gyr, the star formation timescale $\tau$ between 0.03 and 1.0 Gyr, and the stellar metallicity between $Z$ = 0.001-1 $Z_\odot$. The total formed stellar mass is sampled with a logarithmic prior over the range $\log(M_\ast/M_\odot)$ = 5-12.

For dust attenuation, we adopt the flexible attenuation law of \citet{Salim2018ApJ...859...11S}, which extends the \citet{Calzetti2000ApJ...533..682C} prescription by allowing variations in both the slope of the attenuation curve and the strength of the 2175 \AA~bump. Uniform priors are adopted, with the $V$-band attenuation $A_V$ allowed to vary between 0 and 2 mag, the slope parameter $\delta$ between -1.2 and 0.4, and the UV bump strength $B$ between 0 and 2.

For the nebular component, we use a wide prior on the ionization parameter, -4 $<$ log $U$ $<$ -1, consistent with values observed in high-redshift star-forming galaxies and expected for EMPGs. The redshift is fixed to the spectroscopic value of $z = 3.654$.

\begin{figure}
\centering
\includegraphics[width=\linewidth]{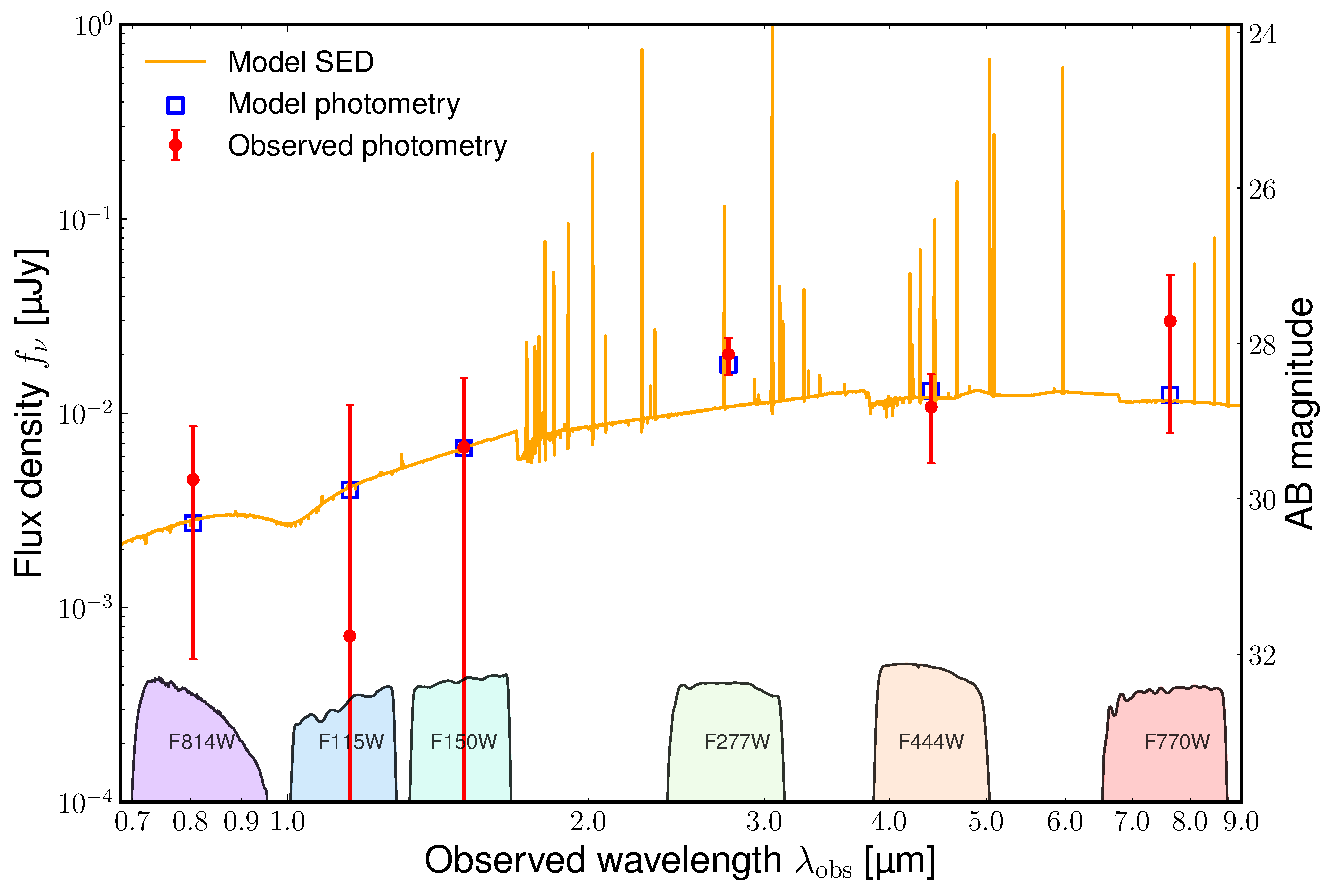} 
\includegraphics[width=\linewidth]{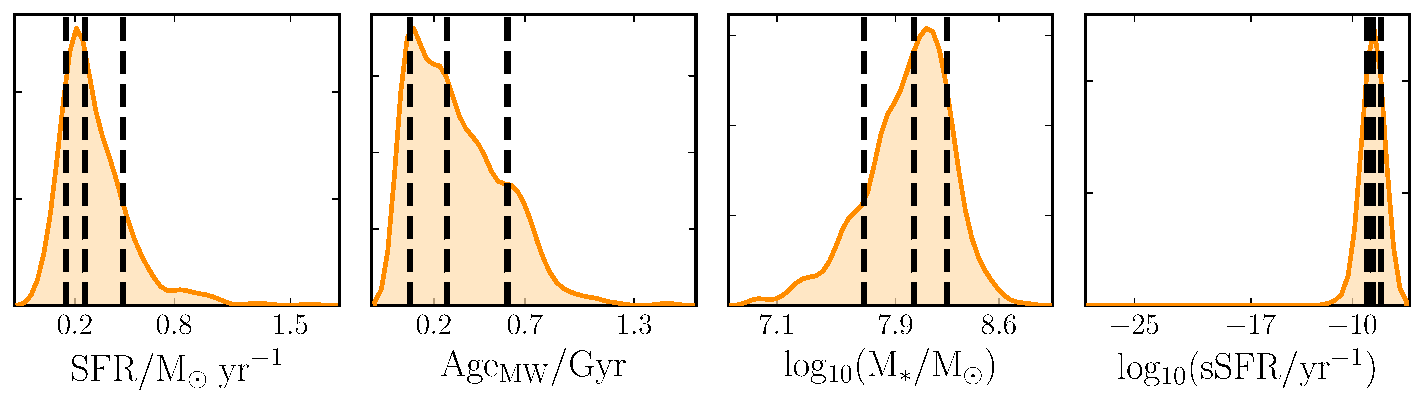} 
\caption{SED modeling of the \target. The upper panel shows the results of SED fitting to the JWST NIRCam photometry. The red points represent the observed photometric measurements, while the orange curve denotes the best-fitting model SED. The blue squares overlaid on the model indicate the synthetic photometry computed from the best-fit model.
The lower panel presents the posterior probability distributions of the SED model parameters. The black dashed lines mark the 16th, 50th, and 84th percentiles of the posterior distributions.}
\label{fig:sed}
\end{figure}

The posterior distributions are sampled using the \texttt{Nautilus} nested sampling algorithm \citep{Lange2023MNRAS.525.3181L}. The results of the SED fitting are shown in Fig.~\ref{fig:sed}. We report the median values of the posterior distributions along with $1\sigma$ uncertainties, corresponding to the 16th and 84th percentiles. The stellar mass is $\mathrm{log}_{10}M_\star/M_\odot = 8.02^{+0.22}_{-0.34}$, the age is $t_\mathrm{age} = 0.27^{+0.33}_{-0.20}$ Gyr, the star formation rate is $\mathrm{SFR} = 0.26^{+0.23}_{-0.12}\,M_\odot\,\mathrm{yr}^{-1}$, corresponding to a specific star formation rate of $\mathrm{log_{10}(sSFR/yr^{-1})} = -8.56^{+0.54}_{-0.43}$. The SED fitting yields a stellar dust attenuation of $A_V = 0.48^{+0.56}_{-0.32}$ mag.

\section{Results and Discussions}
\label{sec:results}

\begin{figure*}
    \centering
    \includegraphics[width=\textwidth]{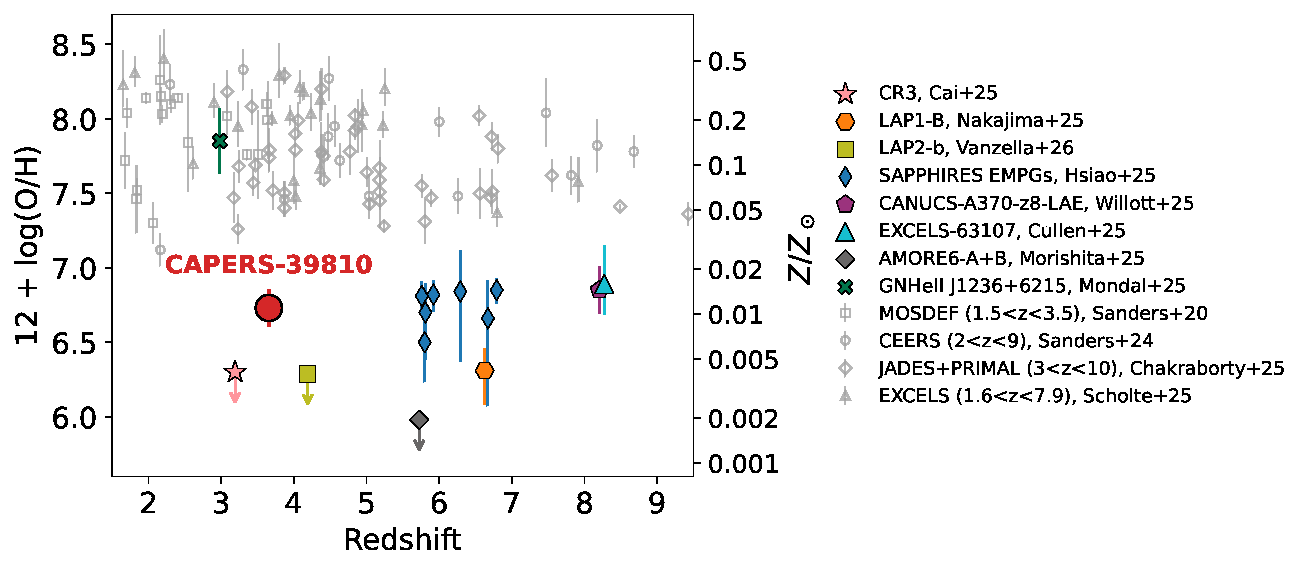} 
    \caption{Gas-phase metallicity of the \target. For comparison, the gray markers in the figure denote the NIRSpec samples for which metallicities have been determined using the direct $T_{e}$ method (\citealt{Sanders2020MNRAS.491.1427S}; \citealt{Sanders2024ApJ...962...24S}; \citealt{Chakraborty2025ApJ...985...24C}; \citealt{Scholte2025MNRAS.540.1800S}), whereas the recently reported metal-poor galaxies are highlighted with colored markers (\citealt{Cai2025ApJ...993L..52C}; \citealt{Nakajima2025arXiv250611846N}; \citealt{Vanzella2026A&A...705L..12V}; \citealt{Hsiao2025arXiv250503873H}; \citealt{Willott2025ApJ...988...26W}; \citealt{Cullen2025MNRAS.540.2176C}; \citealt{Morishita2025arXiv250710521M}; \citealt{Mondal2025ApJ...988..171M}). All upper limits in the figure are 1$\sigma$ values.
}
    \label{fig:Z-z}
\end{figure*}

\begin{figure}
\centering
\includegraphics[width=\linewidth]{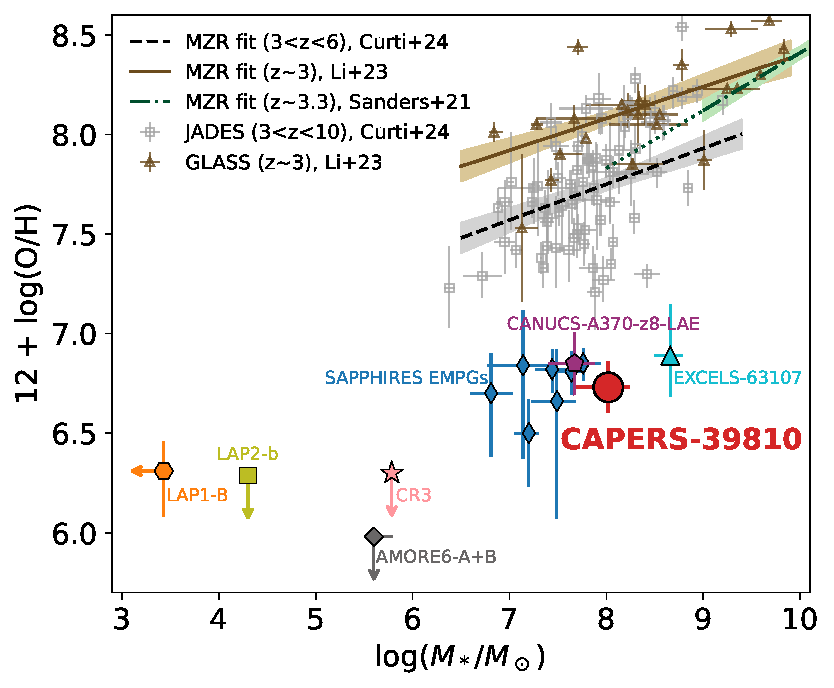} 
\caption{The stellar mass-metallicity relation. Colored symbols denote candidate extremely metal-poor systems, corresponding to the sources highlighted in Fig. \ref{fig:Z-z}. The black dashed, brown solid, and dark-green dot-dashed curves show the best-fit MZR relations at $z\sim3$ from \citet{Curti2024A&A...684A..75C}, \citet{Li2023ApJ...955L..18L}, and \citet{Sanders2021ApJ...914...19S}, respectively, with their 1$\sigma$ uncertainties indicated by the shaded regions; the dark-green dotted lines denote the extrapolation of the \citet{Sanders2021ApJ...914...19S} relation. The overplotted gray squares and brown triangles show galaxy samples from the JWST JADES and GLASS surveys.}
\label{fig:MZR}
\end{figure}

We report a dwarf galaxy, \target, at $z = 3.654$ in the COSMOS field, enabled by the unprecedented depth of the largest NIRSpec MSA program to date, CAPERS. This source adds a new extremely metal-poor galaxy to the intermediate-redshift sample (see Fig.~\ref{fig:Z-z}). Its NIRSpec prism spectrum shows significant detections ($> 3\sigma$) of \hb, \oiii\,$\lambda\lambda$4959, 5007, \ha, and \hei\,$\lambda$10830. The measured line fluxes are summarized in Table~\ref{tab:emission}. 

Based on the SED fitting with \texttt{Bagpipes}, we derive a stellar mass of $\mathrm{log}_{10}M_\star/M_\odot = 8.02^{+0.22}_{-0.34}$, and a stellar population age of $t_\mathrm{age} = 0.27^{+0.33}_{-0.20}$ Gyr for the target. Given its low stellar mass, we adopt the \citet{Sanders2024ApJ...962...24S} R3 calibration and use the mass-metallicity relation to break the degeneracy between the upper and lower metallicity branches. This yields a gas-phase metallicity of $12 + \log(\mathrm{O/H}) = 6.73\pm0.13$. Additionally, using the mass-excitation (MEx) diagram reported by \citet{Coil2015ApJ...801...35C}, which relates R3 to stellar mass, we can rule out active galactic nuclei (AGN) contribution in this system.

The Balmer decrement derived from the observed \ha/\hb\ implies a nebular attenuation of $A_V^{\mathrm{gas}} = 2.7 \pm 1.0$ assuming the \citet{Calzetti2000ApJ...533..682C} attenuation law. This value is larger than the stellar attenuation inferred from the SED fitting (Section~\ref{sec:sed}). 
If interpreted purely as dust reddening, the large Balmer decrement may reflect localized dust associated with recent star-forming regions rather than the global dust content of the galaxy. A patchy or clumpy dust geometry can also produce strong attenuation along particular sightlines while leaving much of the stellar continuum relatively unobscured \citep[e.g.,][]{Wild2011MNRAS.417.1760W}. In addition, the \hb\ line has relatively low SNR in the prism spectrum, and measurement uncertainties may contribute to an elevated \ha/\hb\ ratio. 

Considering the known systematic uncertainties among different metallicity calibrations, it is necessary to test multiple approaches to ensure a robust metallicity estimate. Using the calibration from \citet{Nakajima2022ApJS..262....3N}, whose sample includes a large number of local metal-poor galaxies, we find that our target’s $\mathrm{EW_{0}}(\mathrm{H}\beta)$ lies near the boundary between their “medium-EW’’ and “large-EW’’ subsamples. Applying the medium-EW calibration yields $12 + \log(\mathrm{O/H}) = 7.08 \pm 0.15$, whereas the large-EW calibration gives $12 + \log(\mathrm{O/H}) = 6.95 \pm 0.12$. Although different calibrations introduce systematic offsets of up to $\sim0.3$ dex, the conclusion that our source is an extremely metal-poor galaxy remains unaffected.

\target therefore represents another JWST-confirmed extremely metal-poor galaxy at intermediate redshift, following the recent discoveries reported by \citet{Cai2025ApJ...993L..52C} and \citet{Vanzella2026A&A...705L..12V} (see Fig. \ref{fig:Z-z}). Fig.~\ref{fig:MZR} shows the stellar mass-metallicity relation \citep{Sanders2021ApJ...914...19S,Li2023ApJ...955L..18L,Curti2024A&A...684A..75C}. \target\ lies significantly below the MZR defined by star-forming galaxies at $z\sim3$, making it a pronounced outlier relative to the bulk population. Its detection extends the census of EMPGs to the redshift range between local analogs and the numerous metal-poor systems now uncovered at $z \gtrsim 6$ (\citealt{Hsiao2025arXiv250503873H}; \citealt{Willott2025ApJ...988...26W}; \citealt{Cullen2025MNRAS.540.2176C}; \citealt{Morishita2025arXiv250710521M}; \citealt{Nakajima2025arXiv250611846N}). Such extremely metal-poor systems may in fact be common throughout the Universe. However, current observational constraints make them particularly difficult to identify at intermediate redshifts. The limited survey volume of existing NIRSpec programs, the low SNR at the blue end of the prism spectra, the intrinsically faint nature of EMPGs, and the high observational cost of grating spectroscopy have collectively restricted systematic studies of EMPGs around the cosmic noon. Expanding the census of such galaxies in this epoch is essential for tracing the chemical enrichment history of the Universe, as it directly probes the conditions under which the first generations of metal-poor stars continue to form in low-mass systems.


\normalem
\begin{acknowledgements}
We thank Prof. Zheng Cai's supervision and encouragement of this topics, carefully reading and comment on 
this paper.  He affirmed the contributions of our student team, and decides to remain uncredited. 
 This work is supported by National Key R\&D Program
of China (grant no. 2023YFA1605600), This research is supported by National Natural Science 
Foundation of China (\#12525303)  and Tsinghua
University Initiative Scientific Research Program. This work is based in part on observations made with the NASA/ESA/CSA James Webb Space Telescope. The data were obtained from the Mikulski Archive for Space Telescopes at the Space Telescope Science Institute, which is operated by the Association of Universities for Research in Astronomy, Inc., under NASA contract NAS 5-03127 for JWST. The authors acknowledge the COSMOS-Web team led by CoPIs (J. Kartaltepe and C. Casey) and CAPERS team led by PI (M. Dickinson). All the JWST raw data used in this paper can be found in MAST: \href{https://doi.org/10.17909/fynv-ba75}{10.17909/fynv-ba75}.
\end{acknowledgements}.

\bibliographystyle{raa}
\bibliography{bibtex}

@ARTICLE{Kroupa2001MNRAS.322..231K,
       author = {{Kroupa}, Pavel},
        title = "{On the variation of the initial mass function}",
      journal = {Monthly Notices of the Royal Astronomical Society},
         year = 2001,
        month = apr,
       volume = {322},
       number = {2},
        pages = {231-246},
          doi = {10.1046/j.1365-8711.2001.04022.x},
archivePrefix = {arXiv},
       eprint = {astro-ph/0009005},
 primaryClass = {astro-ph},
       adsurl = {https://ui.adsabs.harvard.edu/abs/2001MNRAS.322..231K}
}

@ARTICLE{Nakajima2023ApJS..269...33N,
       author = {{Nakajima}, Kimihiko and {Ouchi}, Masami and {Isobe}, Yuki and {Harikane}, Yuichi and {Zhang}, Yechi and {Ono}, Yoshiaki and {Umeda}, Hiroya and {Oguri}, Masamune},
        title = "{JWST Census for the Mass-Metallicity Star Formation Relations at z = 4-10 with Self-consistent Flux Calibration and Proper Metallicity Calibrators}",
      journal = {\apjs},
         year = 2023,
        month = dec,
       volume = {269},
       number = {2},
          eid = {33},
        pages = {33},
          doi = {10.3847/1538-4365/acd556},
archivePrefix = {arXiv},
       eprint = {2301.12825},
 primaryClass = {astro-ph.GA},
       adsurl = {https://ui.adsabs.harvard.edu/abs/2023ApJS..269...33N}
}

@ARTICLE{Sarkar2025ApJ...978..136S,
       author = {{Sarkar}, Arnab and {Chakraborty}, Priyanka and {Vogelsberger}, Mark and {McDonald}, Michael and {Torrey}, Paul and {Garcia}, Alex M. and {Khullar}, Gourav and {Ferland}, Gary J. and {Forman}, William and {Wolk}, Scott and {Schneider}, Benjamin and {Bautz}, Mark and {Miller}, Eric and {Grant}, Catherine and {ZuHone}, John},
        title = "{Unveiling the Cosmic Chemistry: Revisiting the Mass{\textendash}Metallicity Relation with JWST/NIRSpec at 4 < z < 10}",
      journal = {\apj},
         year = 2025,
        month = jan,
       volume = {978},
       number = {2},
          eid = {136},
        pages = {136},
          doi = {10.3847/1538-4357/ad8f32},
archivePrefix = {arXiv},
       eprint = {2408.07974},
 primaryClass = {astro-ph.GA},
       adsurl = {https://ui.adsabs.harvard.edu/abs/2025ApJ...978..136S}
}

@ARTICLE{Chemerynska2024ApJ...976L..15C,
       author = {{Chemerynska}, Iryna and {Atek}, Hakim and {Dayal}, Pratika and {Furtak}, Lukas J. and {Feldmann}, Robert and {Greene}, Jenny E. and {Maseda}, Michael V. and {Nanayakkara}, Themiya and {Oesch}, Pascal A. and {Fujimoto}, Seiji and {Labb{\'e}}, Ivo and {Bezanson}, Rachel and {Brammer}, Gabriel and {Cutler}, Sam E. and {Leja}, Joel and {Pan}, Richard and {Price}, Sedona H. and {Wang}, Bingjie and {Weaver}, John R. and {Whitaker}, Katherine E.},
        title = "{The Extreme Low-mass End of the Mass{\textendash}Metallicity Relation at z {\ensuremath{\sim}} 7}",
      journal = {\apjl},
         year = 2024,
        month = nov,
       volume = {976},
       number = {1},
          eid = {L15},
        pages = {L15},
          doi = {10.3847/2041-8213/ad8dc9},
archivePrefix = {arXiv},
       eprint = {2407.17110},
 primaryClass = {astro-ph.GA},
       adsurl = {https://ui.adsabs.harvard.edu/abs/2024ApJ...976L..15C}
}

@ARTICLE{Gutcke2022ApJ...941..120G,
       author = {{Gutcke}, Thales A. and {Pfrommer}, Christoph and {Bryan}, Greg L. and {Pakmor}, R{\"u}diger and {Springel}, Volker and {Naab}, Thorsten},
        title = "{LYRA. III. The Smallest Reionization Survivors}",
      journal = {\apj},
         year = 2022,
        month = dec,
       volume = {941},
       number = {2},
          eid = {120},
        pages = {120},
          doi = {10.3847/1538-4357/aca1b4},
archivePrefix = {arXiv},
       eprint = {2209.03366},
 primaryClass = {astro-ph.GA},
       adsurl = {https://ui.adsabs.harvard.edu/abs/2022ApJ...941..120G}
}

@ARTICLE{Calzetti2000ApJ...533..682C,
       author = {{Calzetti}, Daniela and {Armus}, Lee and {Bohlin}, Ralph C. and {Kinney}, Anne L. and {Koornneef}, Jan and {Storchi-Bergmann}, Thaisa},
        title = "{The Dust Content and Opacity of Actively Star-forming Galaxies}",
      journal = {ApJ},
         year = 2000,
        month = apr,
       volume = {533},
       number = {2},
        pages = {682-695},
          doi = {10.1086/308692},
archivePrefix = {arXiv},
       eprint = {astro-ph/9911459},
 primaryClass = {astro-ph},
       adsurl = {https://ui.adsabs.harvard.edu/abs/2000ApJ...533..682C}
}

@ARTICLE{Casey2023ApJ...954...31C,
       author = {{Casey}, Caitlin M. and {Kartaltepe}, Jeyhan S. and {Drakos}, Nicole E. and {Franco}, Maximilien and {Harish}, Santosh and {Paquereau}, Louise and {Ilbert}, Olivier and {Rose}, Caitlin and {Cox}, Isabella G. and {Nightingale}, James W. and {Robertson}, Brant E. and {Silverman}, John D. and {Koekemoer}, Anton M. and {Massey}, Richard and {McCracken}, Henry Joy and {Rhodes}, Jason and {Akins}, Hollis B. and {Allen}, Natalie and {Amvrosiadis}, Aristeidis and {Arango-Toro}, Rafael C. and {Bagley}, Micaela B. and {Bongiorno}, Angela and {Capak}, Peter L. and {Champagne}, Jaclyn B. and {Chartab}, Nima and {Ch{\'a}vez Ortiz}, {\'O}scar A. and {Chworowsky}, Katherine and {Cooke}, Kevin C. and {Cooper}, Olivia R. and {Darvish}, Behnam and {Ding}, Xuheng and {Faisst}, Andreas L. and {Finkelstein}, Steven L. and {Fujimoto}, Seiji and {Gentile}, Fabrizio and {Gillman}, Steven and {Gould}, Katriona M.~L. and {Gozaliasl}, Ghassem and {Hayward}, Christopher C. and {He}, Qiuhan and {Hemmati}, Shoubaneh and {Hirschmann}, Michaela and {Jahnke}, Knud and {Jin}, Shuowen and {Khostovan}, Ali Ahmad and {Kokorev}, Vasily and {Lambrides}, Erini and {Laigle}, Clotilde and {Larson}, Rebecca L. and {Leung}, Gene C.~K. and {Liu}, Daizhong and {Liaudat}, Tobias and {Long}, Arianna S. and {Magdis}, Georgios and {Mahler}, Guillaume and {Mainieri}, Vincenzo and {Manning}, Sinclaire M. and {Maraston}, Claudia and {Martin}, Crystal L. and {McCleary}, Jacqueline E. and {McKinney}, Jed and {McPartland}, Conor J.~R. and {Mobasher}, Bahram and {Pattnaik}, Rohan and {Renzini}, Alvio and {Rich}, R. Michael and {Sanders}, David B. and {Sattari}, Zahra and {Scognamiglio}, Diana and {Scoville}, Nick and {Sheth}, Kartik and {Shuntov}, Marko and {Sparre}, Martin and {Suzuki}, Tomoko L. and {Talia}, Margherita and {Toft}, Sune and {Trakhtenbrot}, Benny and {Urry}, C. Megan and {Valentino}, Francesco and {Vanderhoof}, Brittany N. and {Vardoulaki}, Eleni and {Weaver}, John R. and {Whitaker}, Katherine E. and {Wilkins}, Stephen M. and {Yang}, Lilan and {Zavala}, Jorge A.},
        title = "{COSMOS-Web: An Overview of the JWST Cosmic Origins Survey}",
      journal = {ApJ},
         year = 2023,
        month = sep,
       volume = {954},
       number = {1},
          eid = {31},
        pages = {31},
          doi = {10.3847/1538-4357/acc2bc},
archivePrefix = {arXiv},
       eprint = {2211.07865},
 primaryClass = {astro-ph.GA},
       adsurl = {https://ui.adsabs.harvard.edu/abs/2023ApJ...954...31C}
}

@ARTICLE{Grogin2011ApJS..197...35G,
       author = {{Grogin}, Norman A. and {Kocevski}, Dale D. and {Faber}, S.~M. and {Ferguson}, Henry C. and {Koekemoer}, Anton M. and {Riess}, Adam G. and {Acquaviva}, Viviana and {Alexander}, David M. and {Almaini}, Omar and {Ashby}, Matthew L.~N. and {Barden}, Marco and {Bell}, Eric F. and {Bournaud}, Fr{\'e}d{\'e}ric and {Brown}, Thomas M. and {Caputi}, Karina I. and {Casertano}, Stefano and {Cassata}, Paolo and {Castellano}, Marco and {Challis}, Peter and {Chary}, Ranga-Ram and {Cheung}, Edmond and {Cirasuolo}, Michele and {Conselice}, Christopher J. and {Roshan Cooray}, Asantha and {Croton}, Darren J. and {Daddi}, Emanuele and {Dahlen}, Tomas and {Dav{\'e}}, Romeel and {de Mello}, Du{\'\i}lia F. and {Dekel}, Avishai and {Dickinson}, Mark and {Dolch}, Timothy and {Donley}, Jennifer L. and {Dunlop}, James S. and {Dutton}, Aaron A. and {Elbaz}, David and {Fazio}, Giovanni G. and {Filippenko}, Alexei V. and {Finkelstein}, Steven L. and {Fontana}, Adriano and {Gardner}, Jonathan P. and {Garnavich}, Peter M. and {Gawiser}, Eric and {Giavalisco}, Mauro and {Grazian}, Andrea and {Guo}, Yicheng and {Hathi}, Nimish P. and {H{\"a}ussler}, Boris and {Hopkins}, Philip F. and {Huang}, Jia-Sheng and {Huang}, Kuang-Han and {Jha}, Saurabh W. and {Kartaltepe}, Jeyhan S. and {Kirshner}, Robert P. and {Koo}, David C. and {Lai}, Kamson and {Lee}, Kyoung-Soo and {Li}, Weidong and {Lotz}, Jennifer M. and {Lucas}, Ray A. and {Madau}, Piero and {McCarthy}, Patrick J. and {McGrath}, Elizabeth J. and {McIntosh}, Daniel H. and {McLure}, Ross J. and {Mobasher}, Bahram and {Moustakas}, Leonidas A. and {Mozena}, Mark and {Nandra}, Kirpal and {Newman}, Jeffrey A. and {Niemi}, Sami-Matias and {Noeske}, Kai G. and {Papovich}, Casey J. and {Pentericci}, Laura and {Pope}, Alexandra and {Primack}, Joel R. and {Rajan}, Abhijith and {Ravindranath}, Swara and {Reddy}, Naveen A. and {Renzini}, Alvio and {Rix}, Hans-Walter and {Robaina}, Aday R. and {Rodney}, Steven A. and {Rosario}, David J. and {Rosati}, Piero and {Salimbeni}, Sara and {Scarlata}, Claudia and {Siana}, Brian and {Simard}, Luc and {Smidt}, Joseph and {Somerville}, Rachel S. and {Spinrad}, Hyron and {Straughn}, Amber N. and {Strolger}, Louis-Gregory and {Telford}, Olivia and {Teplitz}, Harry I. and {Trump}, Jonathan R. and {van der Wel}, Arjen and {Villforth}, Carolin and {Wechsler}, Risa H. and {Weiner}, Benjamin J. and {Wiklind}, Tommy and {Wild}, Vivienne and {Wilson}, Grant and {Wuyts}, Stijn and {Yan}, Hao-Jing and {Yun}, Min S.},
        title = "{CANDELS: The Cosmic Assembly Near-infrared Deep Extragalactic Legacy Survey}",
      journal = {ApJS},
         year = 2011,
        month = dec,
       volume = {197},
       number = {2},
          eid = {35},
        pages = {35},
          doi = {10.1088/0067-0049/197/2/35},
archivePrefix = {arXiv},
       eprint = {1105.3753},
 primaryClass = {astro-ph.CO},
       adsurl = {https://ui.adsabs.harvard.edu/abs/2011ApJS..197...35G}
}

@ARTICLE{Koekemoer2011ApJS..197...36K,
       author = {{Koekemoer}, Anton M. and {Faber}, S.~M. and {Ferguson}, Henry C. and {Grogin}, Norman A. and {Kocevski}, Dale D. and {Koo}, David C. and {Lai}, Kamson and {Lotz}, Jennifer M. and {Lucas}, Ray A. and {McGrath}, Elizabeth J. and {Ogaz}, Sara and {Rajan}, Abhijith and {Riess}, Adam G. and {Rodney}, Steve A. and {Strolger}, Louis and {Casertano}, Stefano and {Castellano}, Marco and {Dahlen}, Tomas and {Dickinson}, Mark and {Dolch}, Timothy and {Fontana}, Adriano and {Giavalisco}, Mauro and {Grazian}, Andrea and {Guo}, Yicheng and {Hathi}, Nimish P. and {Huang}, Kuang-Han and {van der Wel}, Arjen and {Yan}, Hao-Jing and {Acquaviva}, Viviana and {Alexander}, David M. and {Almaini}, Omar and {Ashby}, Matthew L.~N. and {Barden}, Marco and {Bell}, Eric F. and {Bournaud}, Fr{\'e}d{\'e}ric and {Brown}, Thomas M. and {Caputi}, Karina I. and {Cassata}, Paolo and {Challis}, Peter J. and {Chary}, Ranga-Ram and {Cheung}, Edmond and {Cirasuolo}, Michele and {Conselice}, Christopher J. and {Roshan Cooray}, Asantha and {Croton}, Darren J. and {Daddi}, Emanuele and {Dav{\'e}}, Romeel and {de Mello}, Duilia F. and {de Ravel}, Loic and {Dekel}, Avishai and {Donley}, Jennifer L. and {Dunlop}, James S. and {Dutton}, Aaron A. and {Elbaz}, David and {Fazio}, Giovanni G. and {Filippenko}, Alexei V. and {Finkelstein}, Steven L. and {Frazer}, Chris and {Gardner}, Jonathan P. and {Garnavich}, Peter M. and {Gawiser}, Eric and {Gruetzbauch}, Ruth and {Hartley}, Will G. and {H{\"a}ussler}, Boris and {Herrington}, Jessica and {Hopkins}, Philip F. and {Huang}, Jia-Sheng and {Jha}, Saurabh W. and {Johnson}, Andrew and {Kartaltepe}, Jeyhan S. and {Khostovan}, Ali A. and {Kirshner}, Robert P. and {Lani}, Caterina and {Lee}, Kyoung-Soo and {Li}, Weidong and {Madau}, Piero and {McCarthy}, Patrick J. and {McIntosh}, Daniel H. and {McLure}, Ross J. and {McPartland}, Conor and {Mobasher}, Bahram and {Moreira}, Heidi and {Mortlock}, Alice and {Moustakas}, Leonidas A. and {Mozena}, Mark and {Nandra}, Kirpal and {Newman}, Jeffrey A. and {Nielsen}, Jennifer L. and {Niemi}, Sami and {Noeske}, Kai G. and {Papovich}, Casey J. and {Pentericci}, Laura and {Pope}, Alexandra and {Primack}, Joel R. and {Ravindranath}, Swara and {Reddy}, Naveen A. and {Renzini}, Alvio and {Rix}, Hans-Walter and {Robaina}, Aday R. and {Rosario}, David J. and {Rosati}, Piero and {Salimbeni}, Sara and {Scarlata}, Claudia and {Siana}, Brian and {Simard}, Luc and {Smidt}, Joseph and {Snyder}, Diana and {Somerville}, Rachel S. and {Spinrad}, Hyron and {Straughn}, Amber N. and {Telford}, Olivia and {Teplitz}, Harry I. and {Trump}, Jonathan R. and {Vargas}, Carlos and {Villforth}, Carolin and {Wagner}, Cory R. and {Wandro}, Pat and {Wechsler}, Risa H. and {Weiner}, Benjamin J. and {Wiklind}, Tommy and {Wild}, Vivienne and {Wilson}, Grant and {Wuyts}, Stijn and {Yun}, Min S.},
        title = "{CANDELS: The Cosmic Assembly Near-infrared Deep Extragalactic Legacy Survey{\textemdash}The Hubble Space Telescope Observations, Imaging Data Products, and Mosaics}",
      journal = {ApJS},
         year = 2011,
        month = dec,
       volume = {197},
       number = {2},
          eid = {36},
        pages = {36},
          doi = {10.1088/0067-0049/197/2/36},
archivePrefix = {arXiv},
       eprint = {1105.3754},
 primaryClass = {astro-ph.CO},
       adsurl = {https://ui.adsabs.harvard.edu/abs/2011ApJS..197...36K}
}

@ARTICLE{Scoville2007ApJS..172....1S,
       author = {{Scoville}, N. and {Aussel}, H. and {Brusa}, M. and {Capak}, P. and {Carollo}, C.~M. and {Elvis}, M. and {Giavalisco}, M. and {Guzzo}, L. and {Hasinger}, G. and {Impey}, C. and {Kneib}, J. -P. and {LeFevre}, O. and {Lilly}, S.~J. and {Mobasher}, B. and {Renzini}, A. and {Rich}, R.~M. and {Sanders}, D.~B. and {Schinnerer}, E. and {Schminovich}, D. and {Shopbell}, P. and {Taniguchi}, Y. and {Tyson}, N.~D.},
        title = "{The Cosmic Evolution Survey (COSMOS): Overview}",
      journal = {ApJS},
         year = 2007,
        month = sep,
       volume = {172},
       number = {1},
        pages = {1-8},
          doi = {10.1086/516585},
archivePrefix = {arXiv},
       eprint = {astro-ph/0612305},
 primaryClass = {astro-ph},
       adsurl = {https://ui.adsabs.harvard.edu/abs/2007ApJS..172....1S}
}

@ARTICLE{Shuntov2025A&A...704A.339S,
       author = {{Shuntov}, Marko and {Akins}, Hollis B. and {Paquereau}, Louise and {Casey}, Caitlin M. and {Ilbert}, Olivier and {Arango-Toro}, Rafael C. and {McCracken}, Henry Joy and {Franco}, Maximilien and {Harish}, Santosh and {Kartaltepe}, Jeyhan S. and {Koekemoer}, Anton M. and {Yang}, Lilan and {Huertas-Company}, Marc and {Berman}, Edward M. and {McCleary}, Jacqueline E. and {Toft}, Sune and {Gavazzi}, Rapha{\"e}l and {Achenbach}, Mark J. and {Bertin}, Emmanuel and {Brinch}, Malte and {Champagne}, Jackie and {Chartab}, Nima and {Drakos}, Nicole E. and {Egami}, Eiichi and {Endsley}, Ryan and {Faisst}, Andreas L. and {Fan}, Xiaohui and {Flayhart}, Carter and {Hartley}, William G. and {Hatamnia}, Hossein and {Gozaliasl}, Ghassem and {Gentile}, Fabrizio and {Jermann}, Iris and {Jin}, Shuowen and {Kakiichi}, Koki and {Khostovan}, Ali Ahmad and {K{\"u}mmel}, Martin and {Laigle}, Clotilde and {Laishram}, Ronaldo and {Lambrides}, Erini and {Liu}, Daizhong and {Lyu}, Jianwei and {Magdis}, Georgios and {Mobasher}, Bahram and {Moutard}, Thibaud and {Renzini}, Alvio and {Rich}, R. Michael and {Sanders}, David B. and {Sattari}, Zahra and {Robertson}, Brant E. and {Schefer}, Marc and {Scognamiglio}, Diana and {Scoville}, Nick and {Silverman}, John D. and {Taamoli}, Sina and {Trakhtenbrot}, Benny and {Valentino}, Francesco and {Wang}, Feige and {Weaver}, John R. and {Yang}, Jinyi},
        title = "{COSMOS2025: The COSMOS-Web galaxy catalog of photometry, morphology, redshifts, and physical parameters from JWST, HST, and ground-based imaging}",
      journal = {\aap},
         year = 2025,
        month = dec,
       volume = {704},
          eid = {A339},
        pages = {A339},
          doi = {10.1051/0004-6361/202555799},
archivePrefix = {arXiv},
       eprint = {2506.03243},
 primaryClass = {astro-ph.GA},
       adsurl = {https://ui.adsabs.harvard.edu/abs/2025A&A...704A.339S}
}

@ARTICLE{Bromm2011ARA&A..49..373B,
       author = {{Bromm}, Volker and {Yoshida}, Naoki},
        title = "{The First Galaxies}",
      journal = {\araa},
         year = 2011,
        month = sep,
       volume = {49},
       number = {1},
        pages = {373-407},
          doi = {10.1146/annurev-astro-081710-102608},
archivePrefix = {arXiv},
       eprint = {1102.4638},
 primaryClass = {astro-ph.CO},
       adsurl = {https://ui.adsabs.harvard.edu/abs/2011ARA&A..49..373B}
}

@ARTICLE{Schaerer2003A&A...397..527S,
       author = {{Schaerer}, D.},
        title = "{The transition from Population III to normal galaxies: Lyalpha and He II emission and the ionising properties of high redshift starburst galaxies}",
      journal = {\aap},
         year = 2003,
        month = jan,
       volume = {397},
        pages = {527-538},
          doi = {10.1051/0004-6361:20021525},
archivePrefix = {arXiv},
       eprint = {astro-ph/0210462},
 primaryClass = {astro-ph},
       adsurl = {https://ui.adsabs.harvard.edu/abs/2003A&A...397..527S}
}

@ARTICLE{Tumlinson2000ApJ...528L..65T,
       author = {{Tumlinson}, Jason and {Shull}, J. Michael},
        title = "{Zero-Metallicity Stars and the Effects of the First Stars on Reionization}",
      journal = {\apjl},
         year = 2000,
        month = jan,
       volume = {528},
       number = {2},
        pages = {L65-L68},
          doi = {10.1086/312432},
archivePrefix = {arXiv},
       eprint = {astro-ph/9911339},
 primaryClass = {astro-ph},
       adsurl = {https://ui.adsabs.harvard.edu/abs/2000ApJ...528L..65T}
}

@ARTICLE{Wang2024ApJ...967L..42W,
       author = {{Wang}, Xin and {Cheng}, Cheng and {Ge}, Junqiang and {Meng}, Xiao-Lei and {Daddi}, Emanuele and {Yan}, Haojing and {Ji}, Zhiyuan and {Jin}, Yifei and {Jones}, Tucker and {Malkan}, Matthew A. and {Arrabal Haro}, Pablo and {Brammer}, Gabriel and {Oguri}, Masamune and {Hou}, Meicun and {Zhang}, Shiwu},
        title = "{A Strong He II {\ensuremath{\lambda}}1640 Emitter with an Extremely Blue UV Spectral Slope at z = 8.16: Presence of Population III Stars?}",
      journal = {\apjl},
         year = 2024,
        month = jun,
       volume = {967},
       number = {2},
          eid = {L42},
        pages = {L42},
          doi = {10.3847/2041-8213/ad4ced},
archivePrefix = {arXiv},
       eprint = {2212.04476},
 primaryClass = {astro-ph.GA},
       adsurl = {https://ui.adsabs.harvard.edu/abs/2024ApJ...967L..42W}
}

@ARTICLE{Cai2015ApJ...799L..19C,
       author = {{Cai}, Zheng and {Fan}, Xiaohui and {Jiang}, Linhua and {Dav{\'e}}, Romeel and {Oh}, S. Peng and {Yang}, Yujin and {Zabludoff}, Ann},
        title = "{Constraining Very High Mass Population III Stars through He II Emission in Galaxy BDF-521 at z = 7.01}",
      journal = {\apjl},
         year = 2015,
        month = feb,
       volume = {799},
       number = {2},
          eid = {L19},
        pages = {L19},
          doi = {10.1088/2041-8205/799/2/L19},
archivePrefix = {arXiv},
       eprint = {1412.3845},
 primaryClass = {astro-ph.GA},
       adsurl = {https://ui.adsabs.harvard.edu/abs/2015ApJ...799L..19C}
}

@ARTICLE{Cai2011ApJ...736L..28C,
       author = {{Cai}, Zheng and {Fan}, Xiaohui and {Jiang}, Linhua and {Bian}, Fuyan and {McGreer}, Ian and {Dav{\'e}}, Romeel and {Egami}, Eiichi and {Zabludoff}, Ann and {Yang}, Yujin and {Oh}, S. Peng},
        title = "{Probing Population III Stars in Galaxy IOK-1 at z = 6.96 Through He II Emission}",
      journal = {\apjl},
         year = 2011,
        month = aug,
       volume = {736},
       number = {2},
          eid = {L28},
        pages = {L28},
          doi = {10.1088/2041-8205/736/2/L28},
archivePrefix = {arXiv},
       eprint = {1105.2319},
 primaryClass = {astro-ph.CO},
       adsurl = {https://ui.adsabs.harvard.edu/abs/2011ApJ...736L..28C}
}

@ARTICLE{Maiolino2024A&A...687A..67M,
       author = {{Maiolino}, Roberto and {{\"U}bler}, Hannah and {Perna}, Michele and {Scholtz}, Jan and {D'Eugenio}, Francesco and {Witten}, Callum and {Laporte}, Nicolas and {Witstok}, Joris and {Carniani}, Stefano and {Tacchella}, Sandro and {Baker}, William M. and {Arribas}, Santiago and {Nakajima}, Kimihiko and {Eisenstein}, Daniel J. and {Bunker}, Andrew J. and {Charlot}, St{\'e}phane and {Cresci}, Giovanni and {Curti}, Mirko and {Curtis-Lake}, Emma and {de Graaff}, Anna and {Egami}, Eiichi and {Ji}, Zhiyuan and {Johnson}, Benjamin D. and {Kumari}, Nimisha and {Looser}, Tobias J. and {Maseda}, Michael and {Nelson}, Erica and {Robertson}, Brant and {Rodr{\'\i}guez Del Pino}, Bruno and {Sandles}, Lester and {Simmonds}, Charlotte and {Smit}, Renske and {Sun}, Fengwu and {Venturi}, Giacomo and {Williams}, Christina C. and {Willmer}, Christopher N.~A.},
        title = "{JADES. Possible Population III signatures at z = 10.6 in the halo of GN-z11}",
      journal = {\aap},
         year = 2024,
        month = jul,
       volume = {687},
          eid = {A67},
        pages = {A67},
          doi = {10.1051/0004-6361/202347087},
archivePrefix = {arXiv},
       eprint = {2306.00953},
 primaryClass = {astro-ph.GA},
       adsurl = {https://ui.adsabs.harvard.edu/abs/2024A&A...687A..67M}
}

@ARTICLE{Nakajima2025arXiv250611846N,
       author = {{Nakajima}, Kimihiko and {Ouchi}, Masami and {Harikane}, Yuichi and {Vanzella}, Eros and {Ono}, Yoshiaki and {Isobe}, Yuki and {Nishigaki}, Moka and {Tsujimoto}, Takuji and {Nakamura}, Fumitaka and {Xu}, Yi and {Umeda}, Hiroya and {Zhang}, Yechi},
        title = "{An Ultra-Faint, Chemically Primitive Galaxy Forming at the Epoch of Reionization}",
      journal = {arXiv e-prints},
         year = 2025,
        month = jun,
          eid = {arXiv:2506.11846},
        pages = {arXiv:2506.11846},
          doi = {10.48550/arXiv.2506.11846},
archivePrefix = {arXiv},
       eprint = {2506.11846},
 primaryClass = {astro-ph.GA},
       adsurl = {https://ui.adsabs.harvard.edu/abs/2025arXiv250611846N}
}

@ARTICLE{Hsiao2025arXiv250503873H,
       author = {{Hsiao}, Tiger Yu-Yang and {Sun}, Fengwu and {Lin}, Xiaojing and {Coe}, Dan and {Egami}, Eiichi and {Eisenstein}, Daniel J. and {Fudamoto}, Yoshinobu and {Bunker}, Andrew J. and {Fan}, Xiaohui and {Harikane}, Yuichi and {Helton}, Jakob M. and {Kakiichi}, Koki and {Liu}, Yichen and {Liu}, Weizhe and {Maiolino}, Roberto and {Ouchi}, Masami and {Tee}, Wei Leong and {Wang}, Feige and {Wu}, Yunjing and {Xu}, Yi and {Yang}, Jinyi and {Zhu}, Yongda},
        title = "{SAPPHIRES: Extremely Metal-Poor Galaxy Candidates with $12+{\rm log(O/H)}<7.0$ at $z\sim5-7$ from Deep JWST/NIRCam Grism Observations}",
      journal = {arXiv e-prints},
         year = 2025,
        month = may,
          eid = {arXiv:2505.03873},
        pages = {arXiv:2505.03873},
          doi = {10.48550/arXiv.2505.03873},
archivePrefix = {arXiv},
       eprint = {2505.03873},
 primaryClass = {astro-ph.GA},
       adsurl = {https://ui.adsabs.harvard.edu/abs/2025arXiv250503873H}
}

@ARTICLE{Wild2011MNRAS.417.1760W,
       author = {{Wild}, Vivienne and {Charlot}, St{\'e}phane and {Brinchmann}, Jarle and {Heckman}, Timothy and {Vince}, Oliver and {Pacifici}, Camilla and {Chevallard}, Jacopo},
        title = "{Empirical determination of the shape of dust attenuation curves in star-forming galaxies}",
      journal = {\mnras},
         year = 2011,
        month = nov,
       volume = {417},
       number = {3},
        pages = {1760-1786},
          doi = {10.1111/j.1365-2966.2011.19367.x},
archivePrefix = {arXiv},
       eprint = {1106.1646},
 primaryClass = {astro-ph.CO},
       adsurl = {https://ui.adsabs.harvard.edu/abs/2011MNRAS.417.1760W}
}

@ARTICLE{Mondal2025ApJ...988..171M,
       author = {{Mondal}, Chayan and {Saha}, Kanak and {Borgohain}, Anshuman and {Smith}, Brent M. and {Windhorst}, Rogier A. and {Reddy}, Naveen and {Chen}, Chian-Chou and {Umetsu}, Keiichi and {Jansen}, Rolf A.},
        title = "{GNHeII J1236+6215: A He II {\ensuremath{\lambda}}1640 Emitting and Potentially LyC Leaking Galaxy at z = 2.9803 Unveiled through JWST and Keck Observations}",
      journal = {\apj},
         year = 2025,
        month = aug,
       volume = {988},
       number = {2},
          eid = {171},
        pages = {171},
          doi = {10.3847/1538-4357/ade2cd},
archivePrefix = {arXiv},
       eprint = {2506.06831},
 primaryClass = {astro-ph.GA},
       adsurl = {https://ui.adsabs.harvard.edu/abs/2025ApJ...988..171M}
}

@ARTICLE{Sanders2024ApJ...962...24S,
       author = {{Sanders}, Ryan L. and {Shapley}, Alice E. and {Topping}, Michael W. and {Reddy}, Naveen A. and {Brammer}, Gabriel B.},
        title = "{Direct T $_{e}$-based Metallicities of z = 2{\textendash}9 Galaxies with JWST/NIRSpec: Empirical Metallicity Calibrations Applicable from Reionization to Cosmic Noon}",
      journal = {\apj},
         year = 2024,
        month = feb,
       volume = {962},
       number = {1},
          eid = {24},
        pages = {24},
          doi = {10.3847/1538-4357/ad15fc},
archivePrefix = {arXiv},
       eprint = {2303.08149},
 primaryClass = {astro-ph.GA},
       adsurl = {https://ui.adsabs.harvard.edu/abs/2024ApJ...962...24S}
}

@ARTICLE{Chakraborty2025ApJ...985...24C,
       author = {{Chakraborty}, Priyanka and {Sarkar}, Arnab and {Smith}, Randall and {Ferland}, Gary J. and {McDonald}, Michael and {Forman}, William and {Vogelsberger}, Mark and {Torrey}, Paul and {Garcia}, Alex M. and {Bautz}, Mark and {Foster}, Adam and {Miller}, Eric and {Grant}, Catherine},
        title = "{Unveiling the Cosmic Chemistry. II. ``Direct'' T$_{e}$-based Metallicity of Galaxies at 3 < z < 10 with JWST/NIRSpec}",
      journal = {\apj},
         year = 2025,
        month = may,
       volume = {985},
       number = {1},
          eid = {24},
        pages = {24},
          doi = {10.3847/1538-4357/adc7b5},
archivePrefix = {arXiv},
       eprint = {2412.15435},
 primaryClass = {astro-ph.GA},
       adsurl = {https://ui.adsabs.harvard.edu/abs/2025ApJ...985...24C}
}

\end{document}